\documentstyle[12pt]{article}

\begin{document}

\setcounter{page}{1}

\pagestyle{plain} \vspace{1cm}
\begin{center}
\Large{\bf Orbits of particles in noncommutative Schwarzschild spacetime }\\
\small
\vspace{1cm}
{\bf Kourosh Nozari}$^{a,b}$\quad \quad and \quad \quad {\bf Siamak Akhshabi$^{b}$ }\\
\vspace{0.5cm} $^{a}$ {\it Centre for Particle Theory, Durham
University, South Road, Durham DH1 3LE, UK}\\
$^{b}${\it Department of Physics, Faculty of Basic Sciences,
University of Mazandaran,\\
P. O. Box 47416-1467,
Babolsar, IRAN\\
e-mail: kourosh.nozari@durham.ac.uk}

\end{center}
\vspace{1.5cm}
\begin{abstract}
By considering particles as smeared objects, we investigate the
effects of space noncommutativity on the orbits of particles in
Schwarzschild spacetime. The effects of space noncommutativity on
the value of the precession of the perihelion of particle orbit and
deflection of light ray in Schwarzschild geometry are calculated and
the stability of circular orbits is discussed.

\end{abstract}
\newpage
\section{ Introduction}
Gedanken experiments that aim at probing spacetime structure at very
small distances support the idea that noncommutativity of spacetime
is a feature of Planck scale physics. They show that due to
gravitational back reaction, one cannot test spacetime at Planck
scale. Its description as a smooth manifold becomes therefore a
mathematical assumption no more justified by physics. It is then
natural to relax this assumption and conceive a more general
noncommutative spacetime, where uncertainty relations and
discretization naturally arise. Noncommutativity is the central
mathematical concept expressing uncertainty in quantum mechanics,
where it applies to any pair of conjugate variables, such as
position and momentum. One can just as easily imagine that position
measurements might fail to commute and describe this using
noncommutativity of the coordinates. The noncommutativity of
spacetime can be encoded in the commutator[1-7]
\begin{equation}
[\hat{x}^i,\hat{x}^j]=i\theta^{ij}
\end{equation}
where $\theta^{ij}$ is a real, antisymmetric and constant tensor,
which determines the fundamental cell discretization of spacetime
much in the same way as the Planck constant $\hbar$ discretizes the
phase space. In $d = 4$, by a choice of coordinates, this
noncommutativity can be brought to the form
\begin{displaymath}
\theta^{ij}= \left( \begin{array}{cccc}
0 & \theta & 0 & 0  \\
-\theta & 0 & \theta & 0 \\
0 & -\theta & 0 & \theta \\
0 & 0 & -\theta & 0
\end{array} \right)
\end{displaymath}
This was motivated by the need to control the divergences showing up
in theories such as quantum electrodynamics. This noncommutativity
leads to the modification of Heisenberg uncertainty relation in such
a way that prevents one from measuring positions to better
accuracies than the Planck length. In low energy limit, these
quantum gravity effects can be neglected, but in circumstances such
as very early universe or in the strong gravitational field one has
to consider these effects. The purpose of this paper is to
investigate the effects of space noncommutativity on the orbits of a
test particle in noncommutative Schwarzschild geometry. The
modifications induced by the generalized uncertainty principle on
the classical orbits of particles in a central force potential
firstly has been considered by Benczik {\it et al} [8]. The same
problem has been considered within noncommutative geometry by Mirza
and Dehghani[9]and also by Romero and Vergara[10]. The main
consequence of these investigations is the constraint imposed on the
minimal observable length and noncommutativity parameter in
comparison with observational data of Mercury. Recently, stability
of planetary orbits of particles in noncommutative space has been
studied both in central force and Schwarzschild background by Nozari
and Akhshabi[11]. Here we are going to look at the Kepler problem in
noncommutative Schwarzschild geometry from a different viewpoint. It
has been shown that noncommutativity eliminates point-like
structures in favor of smeared objects in flat spacetime[12]. The
effect of smearing can be mathematically implemented as a
substitution rule: position Dirac-delta function can be replaced
everywhere with a Gaussian distribution of minimal width
$\sqrt{\theta}$. In this framework, the mass density of a static,
spherically symmetric, smeared, particle-like gravitational source
can be shown by a Gaussian profile[13]. By adapting such a setup, we
will find a generalized orbit equation. Then we calculate the
modification imposed by space noncommutativity on the value of the
precession of the perihelion of Mercury. The effect of space
noncommutativity on the deflection of light ray in Schwarzschild
geometry is calculated and its numerical value is obtained. Finally
the issue of stability of circular orbits is discussed and the
condition for a circular orbit to be stable in noncommutative
Schwarzschild geometry is obtained.

\section{Noncommutative Schwarzschild Spacetime}
Commutative Schwarzschild geometry is described by the following
line element
\begin{equation}
ds^{2}=\bigg(1-\frac{2M}{r}\bigg)dt^{2}-
\bigg(1-\frac{2M}{r}\bigg)^{-1}dr^{2}-r^{2}d\Omega^{2},
\end{equation}
were we have set $G=1=c$. We want to consider the effect of space
noncommutativity on this line element. Note that it is possible to
consider the effects of space noncommutativity on Einstein field
equations also. One can argue that it is not necessary to change the
Einstein tensor part of the field equations, and that the
noncommutative effects can be implemented acting only on the matter
source[12,13,14]. Since noncommutativity eliminates point-like
objects in favor of smeared objects in flat spacetime, we choose the
mass density of a static, spherically symmetric smeared,
particle-like gravitational source as[13]
\begin{equation}
\rho_{\theta}(r)=\frac{M}{(4\pi\theta)^{\frac{3}{2}}}\exp(-\frac{r^{2}}{4\theta}).
\end{equation}
Solving the Einstein equations with this matter source, one finds
the following noncommutative Schwarzschild line element[13]
\begin{equation}
ds^{2}=\bigg(1-\frac{4M}{r\sqrt{\pi}}\gamma(3/2,r^{2}/4\theta)\bigg)dt^{2}-
\bigg(1-\frac{4M}{r\sqrt{\pi}}\gamma(3/2,r^{2}/4\theta)\bigg)^{-1}dr^{2}-r^{2}d\Omega^{2}
\end{equation}
where $\gamma(3/2,r^{2}/4\theta)$ is the lower incomplete Gamma
function defined as (see appendix)
\begin{equation}
\gamma(3/2,r^{2}/4\theta)\equiv\int^{r^{2/4\theta}}_{0}dt\\\
t^{1/2}e^{-t}
\end{equation}
In the limit of $ r/\sqrt{\theta}\rightarrow\infty$ the classical
Schwarzschild metric is obtained. Using noncommutative Schwarzschild
line element as (4), in the next section we investigate the effect
of space noncommutativity on the orbits of a test particle.
\section{Orbits of Particles}
Along an affinely parameterized geodesic ( timelike , spacelike or
null ) the scalar quantity $ \ 2 K=u^{\alpha}u_{\alpha}$ is a
constant[15]
\begin{equation}
\ 2K=g_{\mu\nu}\frac{dx^{\mu}}{d\lambda}\frac{dx^{\nu}}{d\lambda}.
\end{equation}
If the proper time or proper distance is chosen for $\lambda$, then
$ \ 2K=\pm1$ for timelike and spacelike intervals. For a null
geodesic $ \ 2K=0$. Using the line element (4) we find for the
timelike geodesics
\begin{equation}
2K=\bigg(1-\frac{4M}{r\sqrt{\pi}}\gamma(3/2,r^{2}/4\theta)\bigg)\dot{t}^{2}-
\bigg(1-\frac{4M}{r\sqrt{\pi}}\gamma(3/2,r^{2}/4\theta)\bigg)^{-1}\dot{r}^{2}-
r^{2}\dot{\vartheta}^{2}-r^{2}\sin^{2}(\vartheta)\dot{\varphi}^{2}=1.
\end{equation}
Using the Euler-Lagrange equation
\begin{equation}
\frac{\partial K}{\partial
x^{a}}-\frac{d}{d\tau}\bigg(\frac{\partial K}{\partial
x^{a}}\bigg)=0,
\end{equation}
for $a=0,\, 2,\, 3$ we find
\begin{equation}
\frac{d}{d\tau}\bigg[\Big(1-\frac{4M}{r\sqrt{\pi}}\gamma(3/2,r^{2}/4\theta)\Big)\dot{t}\bigg]=0,
\end{equation}

\begin{equation}
\frac{d}{d\tau}\bigg(r^{2}\dot{\vartheta}-r^{2}\sin\vartheta \cos
\vartheta\dot{\varphi}^{2}\bigg)=0,
\end{equation}
and
\begin{equation}
\frac{d}{d\tau}\bigg(r^{2}sin^{2}\vartheta \dot{\varphi}\bigg)=0.
\end{equation}
These three equations along with equation (7) provide us with the
four equations needed for determining the four desired relations,
namely
$$t=t(\tau),\quad\quad
r=r(\tau),\quad\quad\vartheta=\vartheta(\tau),\quad\quad
\varphi=\varphi(\tau).$$ One also could find the trajectory of the
test particle projected into a slice $t=const.$ To do this end, we
consider motion in the equatorial plane, $\vartheta=\frac{\pi}{2}$.
From the equation (11) we find
\begin{equation}
r^{2}\dot{\varphi}=h,
\end{equation}
where $h$ is constant of integration. Also, integrating equation
(9), we get
\begin{equation}
\bigg(1-\frac{4M}{r\sqrt{\pi}}\gamma(3/2,r^{2}/4\theta)\bigg)\dot{t}=k,
\end{equation}
where $k$ is constant of integration. Substituting (13) into
relation (7), we obtain
\begin{equation}
k^2\bigg(1-\frac{4M}{r\sqrt{\pi}}\gamma\bigg)^{-1}-\bigg(1-\frac{4M}{r\sqrt{\pi}}\gamma\bigg)^{-1}\dot{r}^{2}-r^{2}\dot{\varphi}^{2}=1.
\end{equation}
As usual, we define $u\equiv r^{-1}$ which leads to
\begin{equation}
\dot{r}=-h\frac{du}{d\varphi}.
\end{equation}
Using relations (12), (13) and (15) in equation (14), we find the
following differential equation for the orbit of a test particle in
noncommutative Schwarzschild geometry
\begin{equation}
\bigg(\frac{du}{d\varphi}\bigg)^{2}+u^{2}=\frac{k^{2}-1}{h^{2}}+\frac{4Mu\gamma}{\sqrt{\pi}h^{2}}+\frac{4Mu^{3}\gamma}{\sqrt{\pi}}
\end{equation}
Differentiating (16) we find the second order orbit equation
\begin{equation}
\frac{d^{2}u}{d\varphi^{2}}+u=\frac{2M}{\sqrt{\pi}h^{2}}\gamma-\frac{M}{2\sqrt{\pi}h^{2}\theta^{3/2}}\frac{e^{-1/4\theta
u^{2}}}{u^{3}}+\frac{6Mu^{2}}{\sqrt{\pi}}\gamma-\frac{M}{2\sqrt{\pi}\theta^{3/2}}\frac{e^{-1/4\theta
u^{2}}}{u}.
\end{equation}
This is the orbit equation in noncommutative Schwarzschild
spacetime. If we use the following approximation for incomplete
gamma function (see Appendix)
\begin{equation}
\gamma\bigg(\frac{3}{2},\frac{r^{2}}{4\theta}\bigg)\mid_{\frac{r^{2}}{4\theta}\gg1}\approx\frac{\sqrt{\pi}}{2}+\frac{1}{2}\frac{r}{\sqrt{\theta}}e^{\frac{-r^{2}}{4\theta}}
\end{equation}
then we can write equation (17) as follows
$$\frac{d^{2}u}{d\varphi^{2}}+u=\frac{M}{h^{2}}+3Mu^{2}+\frac{M}{\sqrt{\pi}h^{2}\sqrt{\theta}}\frac{e^{-1/4\theta
u^{2}}}{u}$$
\begin{equation}
-\frac{M}{2\sqrt{\pi}h^{2}\theta^{3/2}}\frac{e^{-1/4\theta
u^{2}}}{u^{3}}+\frac{3Mu}{\sqrt{\pi}\sqrt{\theta}}e^{-1/4\theta
u^{2}}-\frac{M}{2\sqrt{\pi}\theta^{3/2}}\frac{e^{-1/4\theta
u^{2}}}{u}
\end{equation}
The first two terms in the right hand side of equation (19) are the
same as usual general relativity result[16], the other terms are
noncommutative corrections. This equation has a complicated solution
that can be simplified as follows (we set $c=1$)
\begin{equation}
u\simeq\frac{M}{h^{2}}\bigg[1+e\cos
\varphi+3\frac{M^{2}}{h^{2}}e\varphi\sin\varphi
\bigg]+e\frac{M^{2}\sqrt{\theta}}{2\sqrt{\pi}h^{4}}\cos2\varphi+\frac{M^{3}\sqrt{\theta}}{6\sqrt{\pi}h^{5}}e\varphi\sin\varphi
\end{equation}
where
\begin{equation}
e=\frac{Ch^{2}}{M}
\end{equation}
and $C$ is a constant. The last two terms in (20) show the
noncommutative correction and we defined eccentricity  $e$  as in
commutative case. The general form of this result is in agreement
with the one obtained by a different method in reference [9]. Since
noncommutativity parameter is extremely small, the noncommutative
corrections will be very small. However these corrections are
important since they are related to the nature of spacetime
structure at quantum gravity level. Using the relation
$$\cos[(1-\alpha)\varphi]=\cos\varphi+\alpha\frac{d}{d\alpha}\cos[(1-\alpha)\varphi]_{\alpha=0}\\
=\cos\varphi+\alpha\varphi\sin\varphi$$ for small parameter
$\alpha$, one can rewrite equation (20) as follows
\begin{equation}
u\simeq\frac{M}{h^{2}}\Big[1+e\cos(\varphi(1-\alpha))\Big]+e\frac{M^{2}\sqrt{\theta}}{2\sqrt{\pi}h^{4}}\cos2\varphi
\end{equation}
where($c=1$)
\begin{equation}
\alpha=\frac{3M^{2}}{h^{2}}+\frac{M^{2}\sqrt{\theta}}{6\sqrt{\pi}h^{3}}.
\end{equation}
So the period of the orbit is
\begin{equation}
\frac{2\pi}{1-\alpha}\simeq2\pi(1+\alpha)=2\pi(1+\frac{3M^{2}}{h^{2}}+\frac{M^{2}
\sqrt{\theta}}{6\sqrt{\pi}h^{3}}).
\end{equation}
We have therefore found that, during each orbit of the particle,
perihelion advances by an angle
\begin{equation} \Delta
\varphi=2\pi\alpha= \frac{6\pi M^{2}}{h^{2}}+\frac{\pi
M^{2}\sqrt{\theta}}{3\sqrt{\pi}h^{3}}
\end{equation}
This contains an extra precession of the perihelion of the orbit due
to space noncommutativity,
\begin{equation}
\Big(\Delta \varphi\Big)_{NC}=\frac{\pi M^{2}
\sqrt{\theta}}{3\sqrt{\pi}h^{3}}.
\end{equation}
In Newtonian formulation of orbital motion, angular momentum is
given by $h^{2}\approx GM(1-e^{2})a$ \,where $e$ is the eccentricity
and $a$ the semi-major axis of the orbit[15]. Using this relation
and transforming (26) to non-relativistic units we find
\begin{equation}
\Big(\Delta
\varphi\Big)_{NC}=\pi\frac{(GM\theta)^{1/2}}{3c\sqrt{\pi}[(1-e^{2})a]^{3/2}}.
\end{equation}
To have an estimation of this extra precession of perihelion, we
consider the motion of Mercury around Sun. In this case we have
$\frac{GM}{c^2}\simeq 1.48\times 10^{3}\, m$,\, $e=0.2056$,\,
$a=5.79\times 10^{10}\,m$(see for example [15]) and \,$\theta \sim
10^{-72}\, m^{2}$\,[13], we obtain
$$\Big(\Delta \varphi\Big)_{NC}\simeq \pi(0.55\times 10^{-51})\,\,\,radians/orbit=0.355\times 10^{-45}\,\,\, arcseconds/orbit.$$
Mercury orbits once every 88 days, so we find
$$\Big(\Delta \varphi\Big)_{NC}\simeq 0.14722 \times 10^{-42}\,\,\,arcseconds/century.$$

We next consider the case of null geodesics to obtain effect of
space noncommutativity on the light deflection in Schwarzschild
geometry. This can be done by changing the right hand side of
equation (7) to zero
\begin{equation}
2K=\bigg(1-\frac{4M}{r\sqrt{\pi}}\gamma(3/2,r^{2}/4\theta)\bigg)\dot{t}^{2}-
\bigg(1-\frac{4M}{r\sqrt{\pi}}\gamma(3/2,r^{2}/4\theta)\bigg)^{-1}\dot{r}^{2}
-r^{2}\dot{\vartheta}^{2}-r^{2}\sin^{2}(\vartheta)\dot{\varphi}^{2}=0.
\end{equation}
Using the same method as previous part, we reach at the following
second order differential equation for the trajectory of a light ray
in noncommutative Schwarzschild geometry
\begin{equation}
\frac{d^{2}u}{d\varphi^{2}}+u=\frac{6Mu^{2}}{\sqrt{\pi}}\gamma-
\frac{M}{2\sqrt{\pi}\theta^{3/2}}\frac{e^{-1/4\theta u^{2}}}{u}.
\end{equation}
Again using approximation (18) we can rewrite this equation as
\begin{equation}
\frac{d^{2}u}{d\varphi^{2}}+u=3Mu^{2}+\frac{3Mu}{\sqrt{\pi}\sqrt{\theta}}e^{-1/4\theta
u^{2}}- \frac{M}{2\sqrt{\pi}\theta^{3/2}}\frac{e^{-1/4\theta
u^{2}}}{u}
\end{equation}
where the last two terms in the right hand side are the
noncommutative corrections.
 The solution of this equation can be written in the
following form
\begin{equation}
u=u_{0}+u_{1}
\end{equation}
where $u_{0}$ is the special relativistic solution,
\begin{equation}
u_{0}=\frac{1}{D}\sin\varphi,
\end{equation}
and $u_{1}$ has the following form
\begin{equation}
u_{1}\approx
c_{1}\cos\varphi+c_{2}\sin\varphi+\frac{3}{2}\frac{M\sqrt{\theta}[\ln(\sin\varphi)\sin\varphi-\varphi
\cos\varphi]}{D^{3}\sqrt{\pi}}.
\end{equation}
$D$ is a constant with dimension of length. Therefore, the general
solution is
\begin{equation}
u\approx
\frac{\sin\varphi}{D}+c_{1}\cos\varphi+c_{2}\sin\varphi+\frac{3}{2}\frac{M\sqrt{\theta}[\ln(\sin\varphi)\sin\varphi-\varphi
\cos\varphi]}{D^{3}\sqrt{\pi}}
\end{equation}
In a commutative space one can find the solution for $u$ as[16]
\begin{equation}
u\simeq\frac{\sin\varphi}{D}+\frac{M(1+\cos\varphi+\cos^{2}\varphi)}{D^{2}}.
\end{equation}
Therefore we can write $u$ in noncommutative space as follows
\begin{equation}
u\simeq\frac{\sin\varphi}{D}+\frac{M(1+\cos\varphi+\cos^{2}\varphi)}{D^{2}}+\frac{3}{2}\frac{M\sqrt{\theta}[\ln(\sin\varphi)\sin\varphi-\varphi
\cos\varphi]}{D^{3}\sqrt{\pi}}
\end{equation}
The last term in this relation is the correction due to space
noncommutativity. We can now find the angle of deflection for a
light ray using equation (36). If we take the values for $\varphi$
to be $-\epsilon_{1}$ and $\pi+\epsilon_{2}$ when
$r\rightarrow\infty $(see for example [16]),  then using small angle
approximation for $\epsilon_{1} $ and $\epsilon_{2}$, we find the
angle of deflection, $ \delta$, as follows
\begin{equation}
\delta=\epsilon_{1}+\epsilon_{2}=\frac{4M}{D}\bigg[1+\frac{3M\sqrt{\theta}}{2D^{2}\sqrt{\pi}}\bigg]
\end{equation}
Again, the second term is the correction due to space
noncommutativity. Using this relation we try to find a numerical
estimation for noncommutativity effect on the deflection of light
ray. In non-relativistic units the contribution of space
noncommutativity to the angle of deflection becomes
\begin{equation}
\Big(\delta\Big)_{NC}=\frac{6GM}{c^{2}D^{3}}\frac{GM\sqrt{\theta}}{c^{2}\sqrt{\pi}}.
\end{equation}
Since $D= 6.66\times10^{8}m$ (mean radius of sun), for a light ray
just grazing the sun, the approximate value of the noncommutative
correction becomes
$$\Big(\delta\Big)_{NC}\simeq 0.25\times 10^{-55}.$$
Although this correction is very small, but it is important since it
contains information about the nature of spacetime at quantum
gravity level. Note that the value of this term is dependent on the
value of $M$, the mass of central object. So for the case of
very large masses such as very massive stars it may be large enough. \\
From another view point, equation (37) could be used to find a limit
on $\theta$ using observational data of deflection of light rays
around sun.
\section{Stability of Circular Orbits}
For noncommutative Schwarzschild line element as given by equation
(4), there exist two Killing vectors associated with energy and the
angular momentum . In the equatorial plane , $
\vartheta=\frac{\pi}{2} $, the Killing vector associated with energy
is $ \partial_{t}$ or
\begin{equation}
K_{\mu}=\bigg([1-\frac{4M}{r\sqrt{\pi}}\gamma(3/2,r^{2}/4\theta)],\,0,\,0,\,0\bigg)
\end{equation}
and for the angular momentum the Killing vector is $
\partial_{\varphi}$ or
\begin{equation}
L_{\mu}=\Big(0,\,0,\,0,\,-r^{2}sin^{2}\vartheta\Big).
\end{equation}
So, along the geodesics, the two corresponding conserved quantities
are
\begin{equation}
\bigg(1-\frac{4M}{r\sqrt{\pi}}\gamma(3/2,r^{2}/4\theta)\bigg)\frac{dt}{d\lambda}=E
\end{equation}
and
\begin{equation}
r^{2}\frac{d\varphi}{d\lambda}=L
\end{equation}
respectively, where $E$ and $L$ are energy and angular momentum of
the particle per its unit mass. In the equatorial plane, $
\vartheta=\frac{\pi}{2} $ and using equations (9) and (41) we find
\begin{equation}
E^{2}-\bigg(\frac{dr}{d\lambda}\bigg)^{2}+\bigg(1-\frac{4M}{r\sqrt{\pi}}\gamma(3/2,r^{2}/4\theta)\bigg)\bigg(\frac{L^{2}}{r^{2}}+1\bigg)=0
\end{equation}
or
\begin{equation}
\frac{1}{2}\bigg(\frac{dr}{d\lambda}\bigg)^{2}+V(r)=\frac{1}{2}E^{2}
\end{equation}
where we have defined
\begin{equation}
V(r)=\frac{2M}{r\sqrt{\pi}}\gamma(3/2,r^{2}/4\theta)+\frac{2ML^{2}}
{r^{3}\sqrt{\pi}}\gamma(3/2,r^{2}/4\theta)-\frac{L^{2}}{2r^{2}}-\frac{1}{2}
\end{equation}
which is the {\it Effective Potential} in this noncommutative
Schwarzschild spacetime. This noncommutative effective potential has
been shown in figure $1$ in comparison with commutative
Schwarzschild case. The divergency around the origin is a
manifestation of the existence of minimal length scale which
prevents to probe distances smaller than a fundamental distance, for
instance, Planck length. The particle could have a circular orbit at
$ r_{c}$ if
\begin{equation}
\bigg(\frac{\partial V}{\partial r}\bigg)_{r=r_{c}}=0.
\end{equation}
Applying this to the effective potential (45) gives the equation
which determines the radius of circular orbits
\begin{equation}
\frac{-2M}{r^{2}_{c}\sqrt{\pi}}\gamma(3/2,r^{2}/4\theta)+
\frac{r_{c}M}{2\theta^{3/2}\sqrt{\pi}}e^{-r_{c}^{2}/4\theta}
-\frac{6ML^{2}}{r^{4}_{c}\sqrt{\pi}}\gamma(3/2,r^{2}/4\theta)+
\frac{ML^{2}}{2r_{c}\theta^{3/2}\sqrt{\pi}}e^{-r_{c}^{2}/4\theta}
+\frac{L^{2}}{r^{3}_{c}}=0
\end{equation}
so the condition for the stability of the circular orbits is
$$\bigg(\frac{\partial^{2}V}{\partial
r^{2}}\bigg)_{r=r_c}=\frac{4M}{r^{3}_{c}\sqrt{\pi}}\gamma(3/2,r^{2}/4\theta)
+\frac{M}{4\theta^{3/2}\sqrt{\pi}}e^{-r_{c}^{2}/4\theta}
-\frac{M(L^{2}+r_{c}^{2})}{4\theta^{5/2}\sqrt{\pi}}e^{-r_{c}^{2}/4\theta}$$
\begin{equation}
+\frac{24ML^{2}}{r_{c}^{5}\sqrt{\pi}}\gamma(3/2,r^{2}/4\theta)-
\frac{5ML^{2}}{4r_{c}^{2}\theta^{3/2}\sqrt{\pi}}e^{-r_{c}^{2}/4\theta}
-\frac{3L^{2}}{r_{c}^{4}}\geq0
\end{equation}
combining (47) and (48) we find
$$\frac{-2M}{r^{3}_{c}\sqrt{\pi}}\gamma(3/2,r^{2}/4\theta)+
\frac{7M}{4\theta^{3/2}\sqrt{\pi}}e^{-r^{2}_{c}/4\theta}-
\frac{M(L^{2}+r^{2}_{c})}{4\theta^{5/2}\sqrt{\pi}}e^{-r^{2}_{c}/4\theta}$$
\begin{equation}
+\frac{6ML^{2}}{r_{c}^{5}\sqrt{\pi}}\gamma(3/2,r^{2}/4\theta)+
\frac{ML^{2}}{4r^{2}_{c}\theta^{3/2}\sqrt{\pi}}e^{-r^{2}_{c}/4\theta}\geq0
\end{equation}
This is a complicated relation with no analytical solution for
$r_c$. Instead, we have depicted the left hand side of this relation
in terms of radius. The result is shown in figure $2$. In Newtonian
mechanics the circular orbits are stable if $r_{c}\geq 3GM$. In
commutative Schwarzschild geometry this orbits are stable when
$r_{c}\geq 6GM$\,[15]. Now we see that space noncommutativity
increases the radius of stable circular orbits. As figure $5$ shows,
in noncommutative Schwarzschild spacetime the condition for
stability of circular orbits is given by $r_{c}> 6.27\,GM$. So, the
space noncommutativity increases the radius of stable circular
orbits and this is a manifestation of smeared picture of objects in
noncommutative geometry.
\section{Summary and Conclusion}
In this paper we have studied the effects of space noncommutativity
on the orbits of particles in noncommutative Schwarzschild
spacetime. The effects of space noncommutativity is so that it is
not necessary to change the Einstein tensor part of the field
equation, and one can argue that the noncommutative effects can be
implemented acting only on the matter part of Einstein's equations.
Using this picture we have calculated the noncommutative orbital
motion of test particle in noncommutative Schwarzschild spacetime.
An extra precession of the perihelion of the orbit due to space
noncommutativity has been calculated and its numerical value is
estimated using observational data of Mercury. Although this
noncommutative effect is very small, it is important since reflect
the nature of spacetime structure at quantum gravity level. We have
calculated the corrected angle of light deflection due to space
noncommutativity in Schwarzschild spacetime. Noncommutative
effective potential in Schwarzschild spacetime is calculated and its
behavior is compared with commutative result. The stability of
circular orbits in noncommutative Schwarzschild spacetime is
discussed and radius of stable circular orbits is calculated. Radius
of stable circular orbits increases due to space noncommutativity
which is a manifestation of smeared picture of
objects in noncommutative spacetime.\\

\begin{center}
\bf{Appendix}\\ \vspace{0.5cm} {\bf Lower Incomplete Gamma Function}
\end{center}
The lower incomplete gamma function is given by[13]
\begin{equation}
\gamma(a,x)=\int^{x}_{0}t^{a-1}e^{-t}dt=a^{-1}x^{a}e^{-x}\,\,
{^{1}F^{1}}(1;1+a;x)=a^{-1}x^{a}{^{1}F^{1}}(a;1+a;-x)
\end{equation}
where ${^{1}F^{1}}(a;b;x)$is the confluent hypergeometric function
of the first kind. Long and short distance behavior of lower
incomplete gamma function are as follows
\begin{equation}
\gamma\bigg(\frac{3}{2},\frac{r^{2}}{4\theta}\bigg)\mid_{\frac{r^{2}}{4\theta}\ll1}\approx\frac{r^{3}}
{12\sqrt{\theta^{3}}}\bigg(1-\frac{7}{20}\frac{r^{2}}{\theta}\bigg)
\end{equation}
\begin{equation}
\gamma\bigg(\frac{3}{2},\frac{r^{2}}{4\theta}\bigg)\mid_{\frac{r^{2}}{4\theta}\gg1}
\approx\frac{\sqrt{\pi}}{2}+\frac{1}{2}\frac{r}{\sqrt{\theta}}e^{\frac{-r^{2}}{4\theta}}.
\end{equation}

{\bf Acknowledgement}\\
We would like to appreciate professor Jim Bogan for his very helpful
comments on original version of the paper. This work has been done
during KN sabbatical leave at Durham University, UK. He would like
to thank members of the Centre for Particle Theory at Durham
University, specially Professor Ruth Gregory for their hospitality.

\begin{figure}[ht]
\begin{center}
\includegraphics{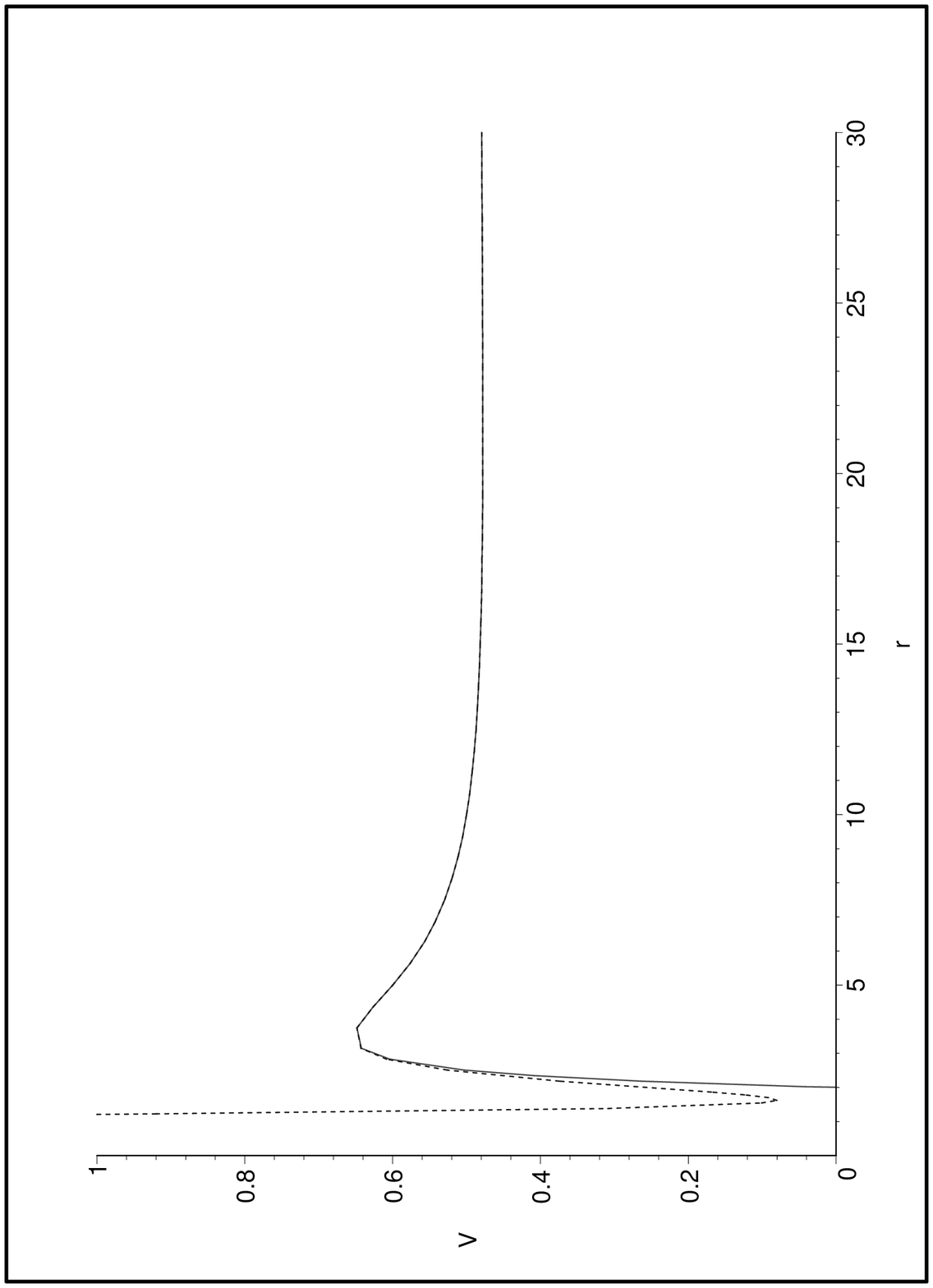}
\end{center}
\vspace{16 cm} \caption{\small {Effective potential in
noncommutative Schwarzschild spacetime(dotted curve) in comparison
with the commutative case(solid line). Divergent behavior of
effective potential around origin in noncommutative case is a
manifestation of existence of minimal length scale which prevents
one to probe distances smaller than this minimal length. This
minimal observable length is of the order of Planck length.}}
 \label{fig:1}
\end{figure}

\begin{figure}[ht]
\begin{center}
\includegraphics{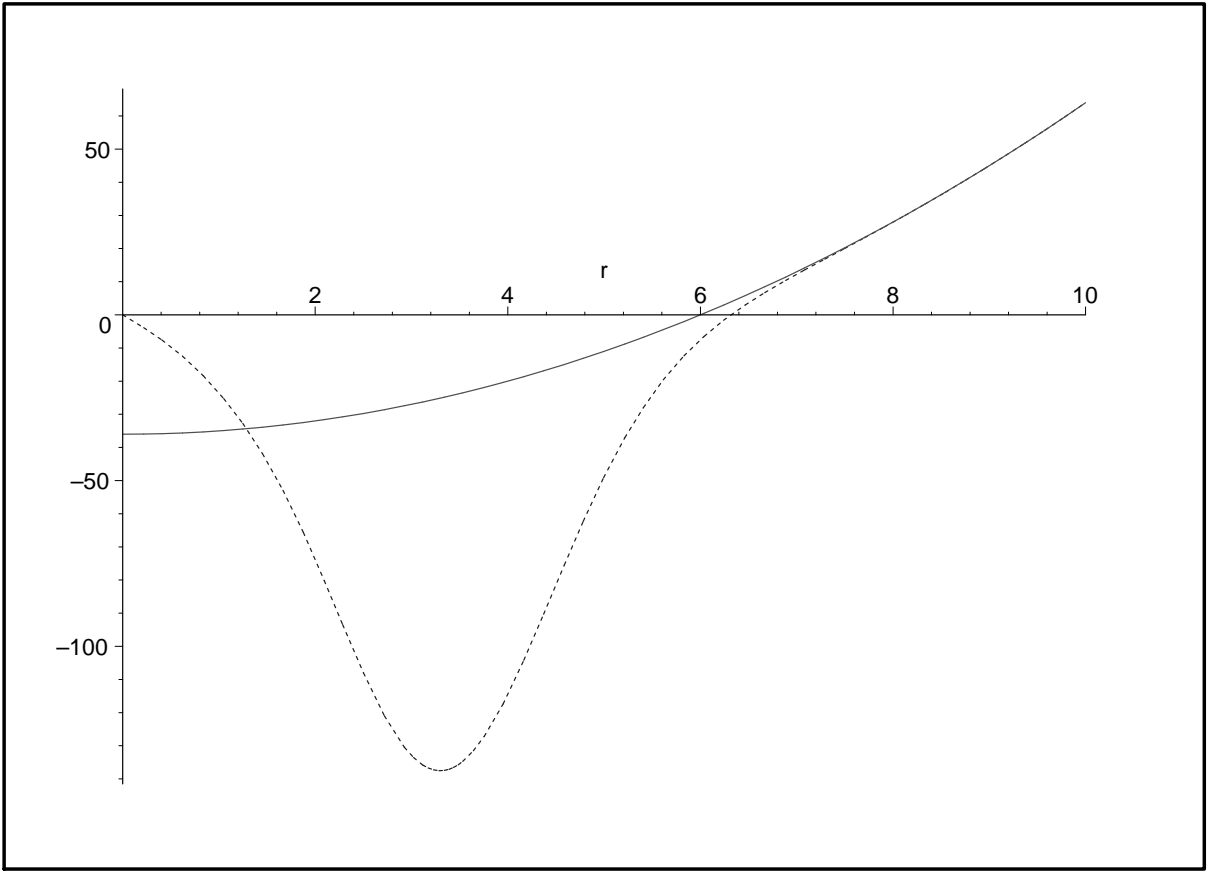}
\end{center}
\vspace{16 cm} \caption{\small {The condition for stability of
circular orbits of particles in Schwarzschild spacetime: In
commutative case the condition for stability is given by $r\geq
6GM$. In noncommutative situation the condition for stability of
circular orbits is given by relation (49) which is shown by
lower(doted) curve. In this case the condition of stability is
$r\geq 6.27\,GM$ }}
 \label{fig:2}
\end{figure}

\end{document}